%% file: Trident.tex
\documentclass[sort&compress,3p,number]{elsarticle}

\usepackage{multirow}
\usepackage{amsmath,amssymb,xcolor}
\usepackage{makecell}
\allowdisplaybreaks

\input{commands}

\begin{document}

\begin{frontmatter}

\title{
Total Born cross section of $e^+e^-$-pair production by an electron in the Coulomb field of a nucleus.
}

\author[binp]{Roman N. Lee}
\ead{r.n.lee@inp.nsk.su}
\author[nsu]{Alexey A. Lyubyakin}
\ead{lyubyakin@inp.nsk.su}
\author[msu,math]{Vladimir A. Smirnov}
\ead{smirnov@theory.sinp.msu.ru}

\address[binp]{Budker Institute of Nuclear Physics, 630090, Novosibirsk, Russia}
\address[nsu]{Novosibirsk State University, 630090, Novosibirsk, Russia}
\address[msu]{Skobeltsyn Institute of Nuclear Physics, Moscow State University, 119992 Moscow, Russia}
\address[math]{Moscow Center for Fundamental and Applied Mathematics, 119992 Moscow, Russia}

\begin{abstract}
We calculate the total Born cross section of the $e^+e^-$-pair production by an electron in the field of a nucleus (trident process) using the modern multiloop methods. For general energies we obtain the cross section in terms of converging power series. The threshold asymptotics and the high-energy asymptotics are obtained analytically. In particular, we obtain additional contribution to the Racah formula due to the identity of the final electrons. Besides, our result for the leading term of the high-energy asymptotics reveals a typo in an old Racah paper \cite{Racah1937}.
\end{abstract}

\end{frontmatter}

\section{Introduction}

The process of the $e^+e^-$-pair production by an electron in the field of a nucleus $e^-Z\to e^-Ze^{+}e^{-}$ (trident process) is one of the basic processes of interaction of high-energy electrons with matter.
Theoretical and experimental studies of this process have a long history \cite{Landau:1934zj,Bhabha:1935pg,Racah1937,PhysRev.88.745,brisbout1956trident,PhysRev.108.1058}.

First theoretical results for the total Born cross section in the leading logarithmic approximation were obtained by Landau and Lifshitz \cite{Landau:1934zj} and, independently, by Bhabha \cite{Bhabha:1935pg}. A little bit later a paper by Racah \cite{Racah1937} appeared, where the total Born cross section of this process was obtained up to power corrections. The Racah result reads
\begin{equation}
\sigma_{e^-Z\to e^-Ze^{+}e^{-}}=\frac{\alpha^{2}(Z\alpha)^{2}}{\pi m^2}
\Biggl(\tfrac{28}{27}L^{3}-\tfrac{178}{27}L^{2}+\left(\tfrac{191}{81}+\tfrac{\pi^{2}}{27}\right)L+\tfrac{683\pi^{2}}{162}-\tfrac{3781}{486}-\tfrac{4\pi^{2}\ln{2}}{9}+\tfrac{79\zeta_3}{9}\Biggr)+\O(1/\gamma)\,.\label{eq:Racah cs}
\end{equation}
Here $L=\ln(2\gamma)$,  $m$ and $\gamma$ are the electron mass and its relativistic factor in the nucleus rest frame. Note that the above formula does not take into account the identity  of the two final electrons. Meanwhile, this account is expected to modify the coefficients in front of $L^1$ and $L^0$ terms.

In the present paper we calculate the total Born cross section of the trident process. We obtain the exact result in terms of convergent power series with analytic coefficients, which allows us to determine analytically both the threshold asymptotics and the high-energy asymptotics of the cross section. In particular, we find that the contribution due to the identity of the final electrons indeed modifies the $\propto L^1$ and $\propto L^0$ terms. We also uncover a typo in the Racah formula which results to the incorrect coefficient $\tfrac{79}{9}$ in front of $\zeta_3$ (the correct one is $\tfrac{185}{18}$).

\section{Details of calculation}

\begin{figure}
\begin{tabular}{c|c}
\includegraphics[scale=0.66]{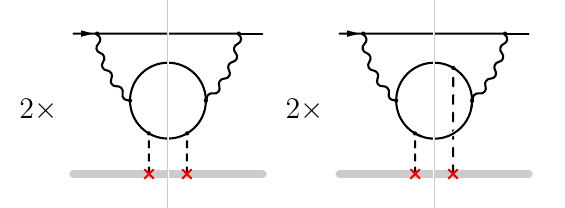} & \includegraphics[scale=0.66]{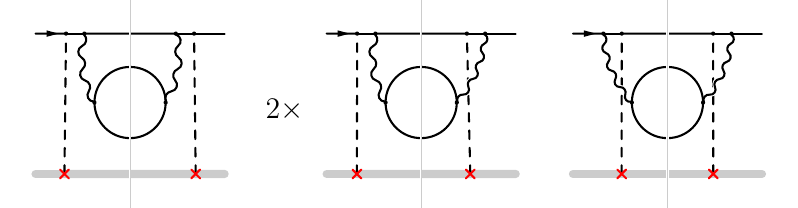}
\\\hline
\multicolumn{2}{c}{\includegraphics[scale=0.66]{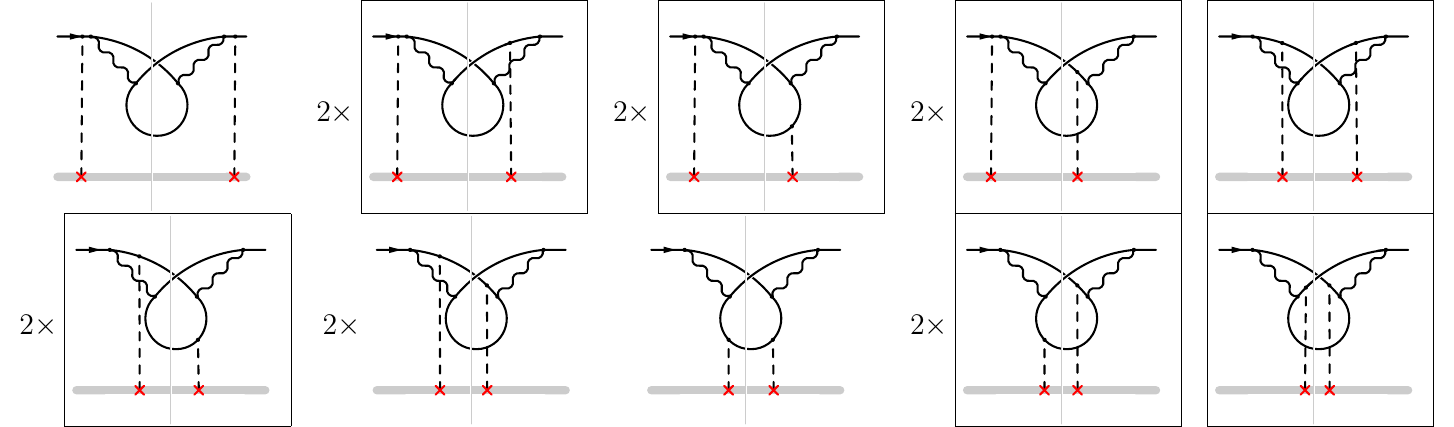}}\\\hline
\end{tabular}
\caption{\label{fig:CS Diagrams}
Diagrams corresponding to the total
Born cross section $Ze+e^-\to Ze+e^-+e^{-}+e^{+}$.
The two upper left diagrams correspond to the $C$-even contribution, the three upper right diagrams correspond to the $C$-odd contribution, and the remaining ten diagrams correspond to the ``twisted'' contribution. Each of the $7$ framed diagrams corresponds to a specific \texttt{LiteRed} basis.}
\end{figure}

The total cross section of the trident process is determined by the three-loop cut diagrams depicted in Fig. \ref{fig:CS Diagrams}. These diagrams fall into three groups:
\begin{itemize}
	\item Diagrams containing cut fermion loop with four photon lines attached.
	\item Diagrams containing cut fermion loop with two photon lines attached.
	\item Diagrams without fermion loop.
\end{itemize}
It is natural to call the corresponding contributions $C$-even, $C$-odd, and ``twisted'' contributions, respectively. The latter contribution appears due to the particle identity.
We will denote these contributions as $\sigma_E$, $\sigma_O$, and $\sigma_T$, so that
\begin{equation}
	\sigma_{e^-Z\to e^-Ze^{-}e^{+}}=\sigma_E + \sigma_O+ \sigma_T\,.
\end{equation}
Let $p_1$, $p_{2,3}$, and $p_4$ denote the momenta of the initial electron, of two final electrons and of the final positron, respectively. Then the cut gray line corresponds to the delta-function $\delta(\ve_1-\ve_2-\ve_3-\ve_3)$, expressing the energy conservation (here $\ve_i=p_i^0 = \sqrt{\boldsymbol{p}_i^2+m^2}$).

We perform Dirac algebra using \texttt{FORM} \cite{Ruijl:2017dtg} and express the three contributions in terms of scalar three-loop integrals with cut denominators.
Each of the three-loop scalar Feynman integrals appearing in the calculation falls into one of the seven families of integrals corresponding to the framed diagrams in Fig. \ref{fig:CS Diagrams}. We perform the IBP reduction using \texttt{LiteRed} and \texttt{FIRE} \cite{Lee2013a,SmirnovSmirnov2013} and identify 74 unique master integrals. We introduce the dimensionless parameter $x=m^2/\ve_1^2=1/\gamma^2$ and derive the differential equation with respect to $x$.

\begin{figure}
	\centering
	\includegraphics[width=0.5\linewidth]{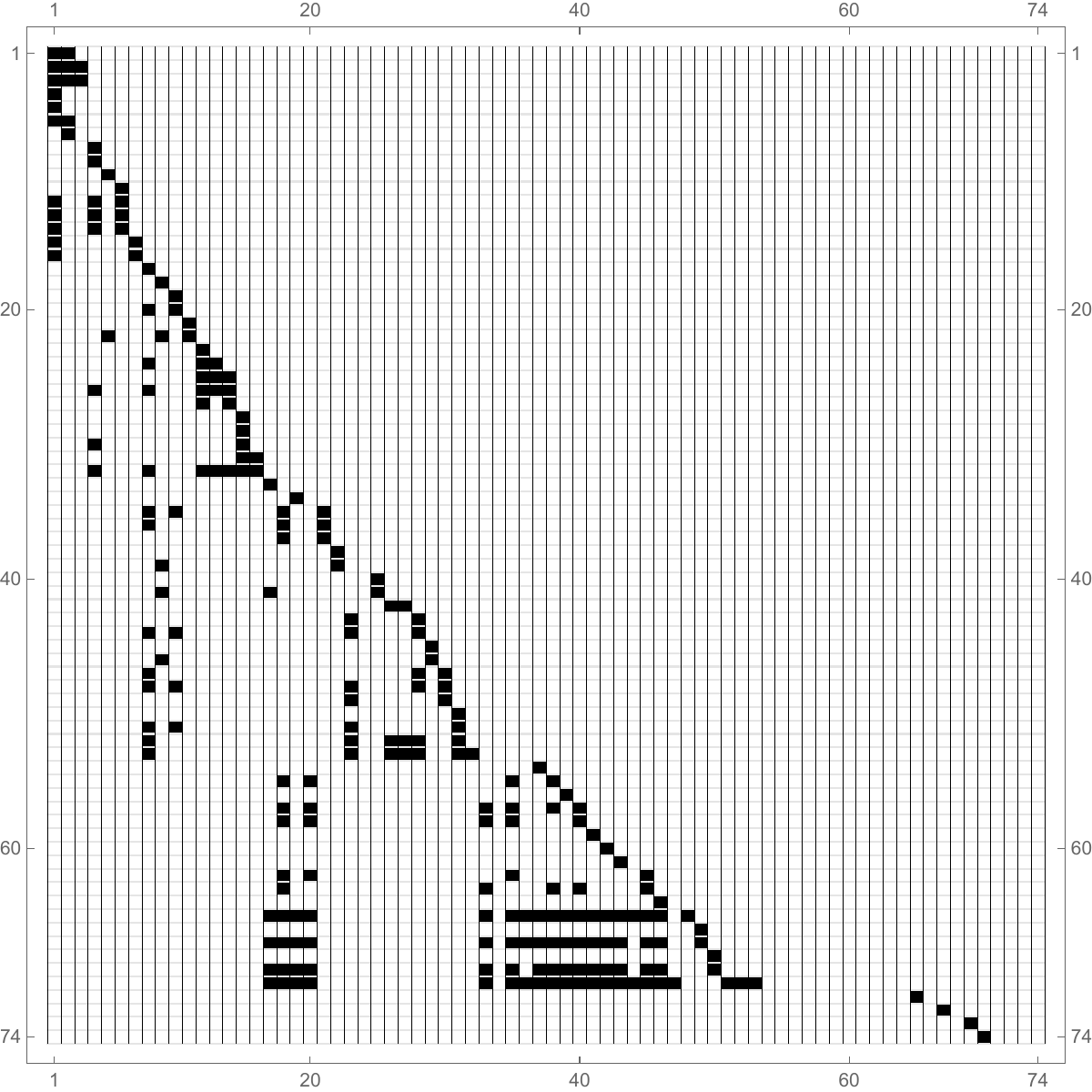}
	\caption{Plot of the matrix on the right-hand side of the differential system for the regular basis. Black squares denote nonzero matrix elements.}
	\label{fig:mreg}
\end{figure}

Using \texttt{Libra} \cite{Lee:2020zfb} and criterion of Ref. \cite{Lee:2017oca}, we have checked that the differential system is irreducible to $\e$-form in a sole $3\times 3$ block, corresponding to the three-particle phase-space integral in the static field,
\begin{equation}
	j_1=\vcenter{\hbox{\includegraphics[scale=0.66]{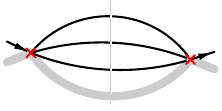}}}=\int \tfrac{d\boldsymbol{p}_2}{2\ve_2}\tfrac{d\boldsymbol{p}_3}{2\ve_3}\tfrac{d\boldsymbol{p}_4}{2\ve_4}\delta(\ve_1-\ve_2-\ve_3-\ve_4)\,,
	\label{eq:j1}
\end{equation}
and its first and second derivatives with respect to $x$. We pass to normalized  Fuchsian form and fix the boundary conditions in the asymptotics $x\to 1/9$, corresponding to the threshold ($\ve_1=3m$). Note that the physical region corresponds to $x\in (0,1/9)$.
The only non-zero boundary constant appears to be the coefficient in the leading asymptotics of $j_1$, Eq. \eqref{eq:j1}, at the threshold. Then we construct an $\e$-regular basis \cite{Lee:2019wwn} following the approach described in Ref. \cite{Krachkov:2023tly}. Once we find the $\e$-regular basis we can safely put $\e=0$ in the differential equations and boundary conditions for it. This is exactly the rationale behind passing to an $\e$-regular basis. When searching for the $\e$-regular basis, it is convenient to pass to the new variable $z$,
\begin{equation}
	x=\left(\tfrac{1-z^{2}}{1+z^{2}}\right)^{2},
	\quad
	z=\sqrt{\tfrac{1-\sqrt{x}}{1+\sqrt{x}}},
	\quad \tfrac1{\sqrt{2}}<z<1\,.
	\label{eq:change var z}
\end{equation}
Then the matrix on the right-hand side of the resulting system at $\e=0$ has many zeros as demonstrated in Fig. \ref{fig:mreg}. In particular, apart from the leftmost upper $3\times 3$ block, all the diagonal elements are zero. The non-zero matrix elements are rational functions of $z$, therefore, we can express all elements of the regular basis as onefold integrals of polylogarithms and the integral $j_1$.

However, the resulting expressions appear to be quite complicated and we choose instead to use the Frobenius method. In order to apply this method, we pass to the variable $\tau$ which varies from $0$ to $\infty$ in the physical region:
\begin{equation}
	x=\left(\tfrac{1-\tau}{3+\tau}\right)^2,\quad \tau=\tfrac{1-3\sqrt{x}}{1+\sqrt{x}},
	\quad 0<\tau<1\,.
\end{equation}
The point $\tau=0$ corresponds to the threshold, while $\tau=1$ corresponds to the high-energy limit.

Then we construct the generalized power series for the evolution operator $U(\tau,\theta)$ around $\theta=0$ and $\theta=1$:
\begin{align}
	U(\tau,\underline{0})&=\sum_{\nu\in\{0,\tfrac32\}}\sum_{n=0}^{N_0-1} U^{(0)}_{n+\nu} \tau^{n+\nu}+ \O(\tau^{N_0+\nu}),\\
	U(\tau,\underline{1})&=\sum_{n=0}^{N_1-1}\sum_{k=0}^7 U^{(1)}_{n,k} \overline{\tau}^n\ln^k \overline{\tau}+  \O(\overline{\tau}^{N_1}),
\end{align}
where $\overline{\tau}=1-\tau$ and the coefficients $U^{(0)}_{n+\nu}$ and $U^{(1)}_{n,k}$ satisfy a finite-order recurrence relation automatically derived using \texttt{Libra}.
Then the specific solution reads
\begin{equation}
	\boldsymbol{J}(\tau)=U(\tau,\underline{0})\boldsymbol{C}_0=U(\tau,\underline{1})\boldsymbol{C}_1\,,
\end{equation}
where $\boldsymbol{C}_0$ and $\boldsymbol{C}_1$ are the two columns of boundary constants. Note that $\boldsymbol{C}_0$ is expressed via the threshold asymptotics of the phase-space integrals $j_1$, which was calculated analytically. Then we can find $\boldsymbol{C}_1$ numerically by matching the two expansions at some point $\tau_0\in (0,1)$:
\begin{equation}
	\boldsymbol{C}_{1}=U^{-1}(\tau_0,\underline{1})U(\tau_0,0)\boldsymbol{C}_{0}.\label{eq:Evolution from threshold to asym}
\end{equation}
Our goal was to obtain high-precision values for $\boldsymbol{C}_{1}$ in order to recognize their analytic form using PSLQ algorithm \cite{FergBai1991}. In order to evaluate $U(\tau_0,\underline{1})$  and $U(\tau_0,\underline{0})$ with comparable precision, we have to choose $N_1$ and $N_0$ so that
\begin{equation}
	\tau_0^{N_0}\approx(1-\tau_0)^{N_1}\,.
\end{equation}
Since the expansion of $U(\tau,\underline{1})$ appears to be much more computationally expensive than that of $U(\tau,\underline{0})$, we choose $\tau_0=9/10$, $N_0=6700$,  $N_1=307$. This choice gives us about $300$ digits of $\boldsymbol{C}_{1}$ within about 10 minutes of computation time for each $U(\tau_0,\underline{1})$  and $U(\tau_0,\underline{0})$.
Then we use PSLQ to obtain the analytic form of $\boldsymbol{C}_{1}$ in terms of alternating multiple zeta values.

\section{Results and Conclusion}

The results obtained within the described approach are the following:
\begin{itemize}
	\item Threshold asymptotics with analytic coefficients to arbitrary order.
	\item High-energy asymptotics  with analytic coefficients to arbitrary order.
	\item High-precision numerical results for arbitrary energies.
\end{itemize}
The threshold asymptotics of $\sigma_E$, $\sigma_O$, and $\sigma_T$ read ($\tau=\tfrac{\gamma-3}{\gamma+1}$)
\begin{align}
	\sigma_E & =\frac{\alpha^{2}(Z\alpha)^{2}}{m^2}\left[\frac{304 \tau ^{9/2}}{945}-\frac{64 \tau ^{11/2}}{10395}+\frac{15040 \tau ^{13/2}}{27027}-\frac{9472 \tau ^{15/2}}{405405}+\O\left(\tau^{17/2}\right)\right],\label{eq:sigEth}\\
	\sigma_O & =\frac{\alpha^{2}(Z\alpha)^{2}}{m^2}\left[
	\frac{8 \tau ^{7/2}}{35}-\frac{232 \tau ^{9/2}}{945}+\frac{512 \tau ^{11/2}}{1485}-\frac{1568 \tau ^{13/2}}{6435}+\frac{11528 \tau ^{15/2}}{61425}
	+\O\left(\tau^{17/2}\right)
	\right],\label{eq:sigOth}\\
	\sigma_T & =\frac{\alpha^{2}(Z\alpha)^{2}}{m^2}\left[
	-\frac{4 \tau ^{7/2}}{21}+\frac{76 \tau ^{9/2}}{135}-\frac{1856 \tau ^{11/2}}{1155}+\frac{395632 \tau ^{13/2}}{135135}-\frac{121276 \tau ^{15/2}}{25025}
	+\O\left(\tau^{17/2}\right)
	\right]\,.\label{eq:sigTth}
\end{align}

The high-energy asymptotics reads
\begin{multline}
	\sigma_E =\frac{\alpha^{2}(Z\alpha)^{2}}{\pi m^2}\bigg\{
	\underline{\frac{28 }{27}L^3-\frac{178}{27} L^2+\left(\frac{430}{27}-\frac{25 \pi ^2}{18}\right) L+\frac{68 \zeta_3}{3}+\frac{13}{9} \pi ^2 \ln2+\frac{877 \pi ^2}{324}-\frac{512}{27}}
	+\frac1{\gamma}\bigg[-\frac{27\pi ^2}{2} L\\
	+36 \pi ^2 \ln2-\frac{\pi ^2}{4}\bigg]
	+\frac1{\gamma^2}\bigg[\frac{L^5}{3}-\frac{5 L^4}{6}+\left(\frac{179}{27}-\frac{5 \pi ^2}{9}\right) L^3
	+\left(4 \zeta_3-\frac{655}{54}+\frac{7 \pi ^2}{6}\right)L^2
	+\bigg(\frac{5 \pi ^4}{18}-16 \zeta_3\\
	-\frac{25 \pi ^2}{2}-\frac{3907}{54}\bigg)L
	-\frac{47 \zeta_5}{2}-\frac{11 \pi ^2 \zeta_3}{6}+\frac{476 \zeta_3}{3}+\frac{\pi ^4}{24}-\frac{286 \pi ^2}{81}-\frac{15161}{108}+\frac{131}{9} \pi ^2 \ln2
	\bigg]
	+\O\left(\tfrac1{\gamma^3}\right)
	\bigg\}\,,\label{eq:sigEhe}
\end{multline}
\begin{multline}
	\sigma_O = \frac{\alpha^{2}(Z\alpha)^{2}}{\pi m^2}\bigg\{
	\underline{\left(\frac{77 \pi ^2}{54}-\frac{1099}{81}\right) L-\frac{223 \zeta_3}{18}+\frac{163 \pi ^2}{108}+\frac{5435}{486}-\frac{17}{9} \pi ^2 \ln2}
	+\frac1\gamma\frac{3 \pi ^2}{4}
	+\frac1{\gamma^2}\bigg[-\frac{7 L^4}{18}+\frac{11 L^3}{9}\\
	+\left(\frac{5 \pi ^2}{18}-\frac{415}{54}\right) L^2+\left(\frac{901}{162}-\frac{\pi ^2}{18}\right) L-\frac{13 \zeta_3}{2}-\frac{17 \pi ^4}{360}+\frac{143 \pi ^2}{324}-\frac{3935}{972}-\frac{2}{9} \pi ^2 \ln2\bigg]
	+\O\left(\tfrac1{\gamma^3}\right)
	\bigg\}\,,\label{eq:sigOhe}
\end{multline}
\begin{multline}
	\sigma_T = \frac{\alpha^{2}(Z\alpha)^{2}}{\pi m^2}\bigg\{
	\underline{\left(-\frac{748 \zeta_3}{105}-\frac{2729}{630}+\frac{13591 \pi ^2}{4725}-\frac{16}{7} \pi ^2 \ln2\right)L} +\frac{93 \zeta_5}{8}-\frac{7 \pi ^2 \zeta_3}{8}-\frac{2048 \text{Li}_4\left(\frac{1}{2}\right)}{35}+\frac{101 \pi ^4}{105}\\
	-\frac{6051 \zeta_3}{175}+\frac{3007 \pi ^2}{4725}-\frac{5282}{1575}-\frac{256 \ln^4{2}}{105}+\frac{496}{105} \pi ^2 \ln^2{2}-\frac{1242}{175} \pi ^2 \ln2
	+\frac1\gamma \frac{27 \pi ^2}{8}
	+\frac1{\gamma^2}\bigg[\frac{L^5}{15}-\frac{5 L^4}{4}+\frac{5 L^3}{18}
	\\
	+ \left(-2 \zeta_3-\frac{91}{4}+\frac{11 \pi ^2}{36}\right)L^2+\left(\frac{88 \zeta_3}{21}-\frac{493}{126}+\frac{6707 \pi ^2}{3780}+\frac{\pi ^4}{15}-\frac{166}{105} \pi ^2 \ln2\right)L -\frac{77 \zeta_5}{4}
	+\frac{11 \pi ^2 \zeta_3}{12}
	\\
	-\frac{1144 \text{Li}_4\left(\frac{1}{2}\right)}{35}
	+\frac{9809 \pi ^4}{15120}+\frac{344\pi ^2}{105}  \ln^2{2}-\frac{143 \ln^4{2}}{105}
	+\frac{257 \zeta_3}{140}-\frac{434 \pi ^2}{45} \ln2+\frac{2836 \pi ^2}{4725}-\frac{746}{21}\bigg]
	+\O\left(\tfrac1{\gamma^3}\right)
	\bigg\}\,.\label{eq:sigThe}
\end{multline}

\begin{figure}
	\centering
	\includegraphics[width=0.30\linewidth]{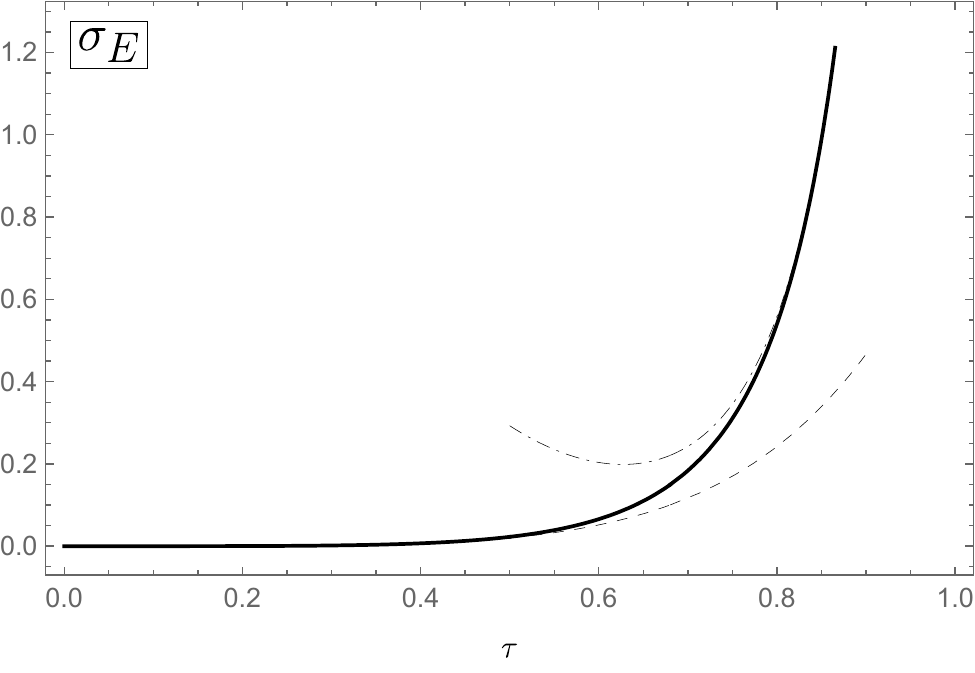}
	\includegraphics[width=0.30\linewidth]{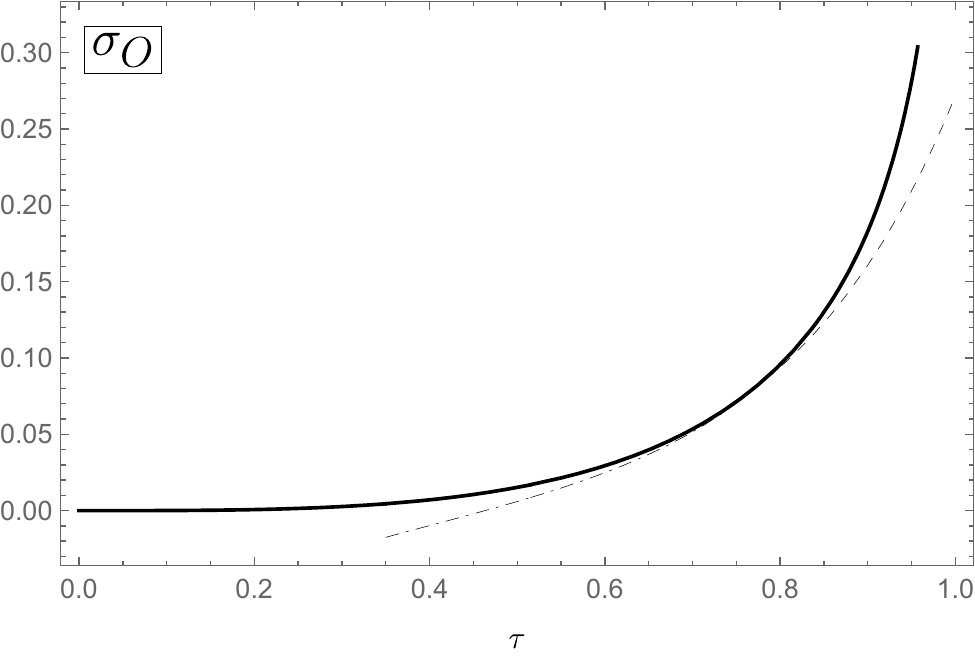}
	\includegraphics[width=0.30\linewidth]{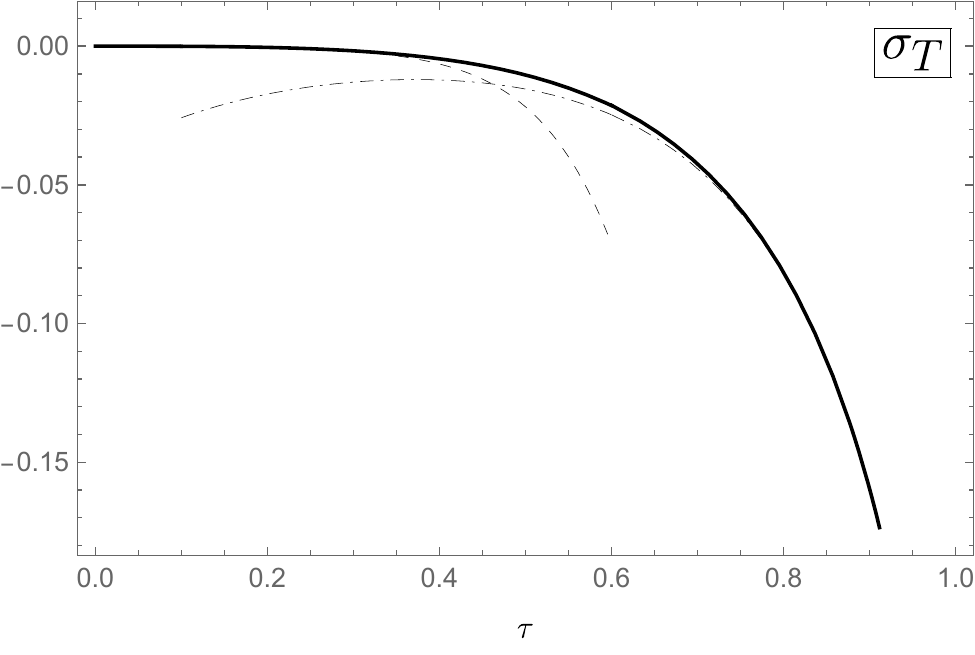}
	\caption{Cross sections  $\sigma_E,\ \sigma_O,\ \sigma_T$ as functions of $\tau$. Dashed and dash-dotted curves correspond to the threshold, Eqs. \eqref{eq:sigEth}-\eqref{eq:sigTth} and high-energy, Eqs. \eqref{eq:sigEhe}-\eqref{eq:sigThe}, asymptotics, respectively.}
	\label{fig:plotEOT}
\end{figure}

Finally, the functions $\sigma_E,\ \sigma_O,\ \sigma_T$ for arbitrary $\tau=\tfrac{\gamma-3}{\gamma+1}$ are presented in Fig.\ref{fig:plotEOT}. The solid graphs were obtained using deep Frobenius expansions near $\tau=0$ and $\tau=1$.

We note that the leading terms underlined in Eqs. \eqref{eq:sigEhe} and \eqref{eq:sigOhe} were also considered by Racah in Ref. \cite{Racah1937}. While we find agreement of Eq. \eqref{eq:sigEhe} with the sum of Eqs. (59) and (66) of Ref. \cite{Racah1937}, the underlined terms in Eq.\eqref{eq:sigOhe} slightly differ from Eq. (70) of Ref. \cite{Racah1937}. Namely, the coefficient in front of $\zeta_3$ is $-\tfrac{125}{9}$ in Ref. \cite{Racah1937}, which is to be compared with $-\tfrac{223}{18}$ in Eq. \eqref{eq:sigOhe}. Fortunately, it is easy to identify the place where this typo appeared: simply integrating Eq. (69) of Ref. \cite{Racah1937} over $dt/t$ we recover our result. It turns out that the logarithmically amplified term in the leading asymptotics of the contribution $\sigma_T$, underlined in Eq. \eqref{eq:sigThe} also can be found in old papers. Namely, Kuraev, Lipatov and Strikman in Ref. \cite{kuraev1973identity} considered the effect of electron identity in the process $e^+e^-\to e^+e^+e^-e^-$. After taking into account the combinatorial factor $2$ due to the appearance of two pairs of identical particles in this process instead of one pair in $e^-Z\to e^-Ze^{-}e^{+}$ we find agreement with Ref. \cite{kuraev1973identity}.

Finally, we note that the leading high-energy asymptotics of the cross sections suffice only for rather high energies. In particular, they provide $5$\% accuracy at $\ve_1\gtrsim 200 m$, to be compared with $\ve_1\gtrsim 50m$ and $\ve_1\gtrsim 17m$ when one takes into account the $1/\gamma$ and $1/\gamma^2$ corrections, respectively.

\paragraph*{Acknowledgments} We are grateful to V. Fadin for useful discussions and interest to the work.
The work of R.L. and A.L. was supported by the Russian Science Foundation, grant number 20-12-00205. The work of V.S. was supported by the Ministry of Education and Science of the Russian Federation as part of the program of the Moscow Center for Fundamental and Applied Mathematics under Agreement No.\ 075-15-2022-284.

\bibliographystyle{elsarticle-num}
\bibliography{Trident.bib}

\end{document}

%% file: commands.tex
\renewcommand{\O}{\mathcal{O}}

\newcommand{\e}{\epsilon}
\newcommand{\ve}{\varepsilon}

\newcommand{\citelater}[1][]{%
	{\color{red}
		\ifthenelse{\equal{#1}{}}{[\raisebox{0.1em}{\tiny \parbox[c]{3.5em}{ insert \\ citation}}]}{[#1]}}
	\xspace%
}